\documentclass[preprint,showpacs,preprintnumbers,amsmath,amssymb]{revtex4}
\usepackage{epsfig}
\usepackage{graphicx}
\usepackage{bm}
\begin{document}

\title{Coupled criticality analysis of inflation and unemployment

}
\author{Z. Koohi Lai$^{1}$, A. Namaki$^{2,*}$, A. Hosseiny$^{3}$, G.R. Jafari$^{3,4}$ , M. Ausloos$^{5,6,7}$\\
{\small $^1$ Department of Physics, Islamic Azad University, Firoozkooh Branch, Firoozkooh, Iran} \\
{\small $^2$ Department of Finance, Faculty of Management, University of Tehran, Tehran, Iran }\\
{\small $^3$ Department of Physics, Shahid Beheshti University, G.C., Evin, Tehran 19839, Iran } \\
{\small $^4$ Center for Network Science, Central European University, 1051 Budapest, Hungary } \\
{\small $^5$ School of Business, College of Social Sciences, Arts, and Humanities, Brookfield, University of Leicester, Leicester, LE2 1RQ, United Kingdom}\\
{\small $^6$ Group of Researchers for Applications of Physics in Economy and Sociology (GRAPES), Rue de la belle jardinière, 483, Sart Tilman, B-4031 Angleur, Liege, Belgium } \\ 
{\small $^7$ Department of Statistics and Econometrics,  Bucharest University of Economic Studies, \\ Calea Dorobantilor 15-17, Bucharest, 010552 Sector 1, Romania. } \\ 
{\small $^*$ Corresponding author, Email: Alinamaki@ut.ac.ir }\\
}

\vskip 1cm

\begin{abstract}

In this paper, we are interested to focus on the critical periods in economy which are characterized by large fluctuations in macroeconomic indicators.
 To capture unusual and large fluctuations of inflation and unemployment, we concentrate on the non-Gaussianity of their distributions.
 To this aim,  by using the coupled multifractal approach, we analyze US data for a period of 70 years from 1948 until 2018 and measure the non-Gausianity of the distributions.Then, we investigate how the non-Gaussianity of the variables affect the coupling structure of them.

By applying the multifractal method,  one can see that  the non-Gaussianity depends on the scales. While the non-Gaussianity of unemployment is noticeable only for periods smaller than 1 year and for longer periods tends to Gaussian behavior, the non-Gaussianities of  inflation  persist for all time scales.
Also, it is observed that the coupling structure of these variables tends to a Gaussian behavior after $2$ years.

\end{abstract}
\maketitle

\section{Introduction}

Unemployment and inflation are two important economic indices; their
relation is meaningful for policy makers. Historically, there have
been hot debates over the relation between unemployment and
inflation,  -  starting with the observation of Phillips
\cite{phillips}.

 Later, the huge debate was ignited about the strength and causal effect of the
relation \cite{solow,friedman,phelps}.  The matter has been  important  because of its influence on    government
policies  during the times of recession, but at other times as well \cite{taylor,mishkin}. Discussion over the matter is still a serious
controversial issue in economics, see for example \cite{akerlof,gali,mankiw}.

Yet, few works have considered the theoretical complexity of these variables,  thus leading to
 the research questions of this paper \cite{safdari}.

Of course, economics  complexity  has attracted a big deal of attention  in recent years \cite{arthur,deligatti,Namaki1,farmer,Hosseiny,namaki,helbing2013rethinking}.

 Economy  can be considered as a big network of heterogenous agents which interact with each other and with their environment  \cite{{Schweitzer}, {d2016complex},{chakrabarti},{Namaki2},{Dragulescu},{Vandewalle},{hosseinyrole},{Aoyama2010},{VRAC015complex},{hosseinygeommetry}}.
It  is reasonable to expect that inflation and unemployment as outcomes of these complex systems inherit complexity considerations.

This suggests a thorough analysis of these variables along advanced techniques available in complexity theory approach and applications \cite{Anderson99,Ivanova2002,Ausloos12}.

 Looking at  unemployment and inflation  indices as simple variables may
ignore some rich knowledge about their complexity. It has been shown
for example that  economic indicators and their coupling have nice
scaling features \cite{Vandewalle98,Vandewalle1998,safdari}. In fact, many economic variables  present some multifractal nature \cite{Ivanova99,Grech,Zhou}.  Such an
observation suggests that inflation and unemployment data sets might
exhibit non-Gaussian probability density function (PDF). This
behavior may originate from the occurrence of large fluctuations in the
system at extreme values.

In order to model the non-Gaussianity of  some signal, a multifractal
random walk (MRW) model  has already been implemented; MRW is composed of the
product of two processes with normal and log-normal PDFs
\cite{Bacry01,Zahra}. The variance of the log-normal part determines the strength
 of non-Gaussianity of the original signal \cite{Chabaud,Ghas96,Ausloos2002}. 

In order to go further, i.e. to capture the non-Gaussianity in the {\it  coupling of  variables}, we apply the  bivariate Multifractal Random Walk (Bi-MRW) technique \cite{Muzy01,Muzy02}. 
 Therefore, this approach leads us to obtain and to analyze  the type of cross-correlations between large fluctuations in inflation and unemployment measures. Whence, the paper is organized as follows. In section 2, we explain the MRW and Bi-MRW methods;  in section 3 and section 4, we present our findings and conclusions respectively.

\section{Methods}

\subsection{Multifractal Random Walk (MRW)}

The Multifractal Random Walk (MRW)  for analysing time series stems from turbulent cascade models \cite{Chabaud}. These processes are useful for presenting the non-Gaussian behavior of time series.

 The temporal fluctuations  increment of a process at scale $s$ ,
$\delta_sx(t)=x(t+s)-x(t)$, can be modeled by the product of a
normal and a log-normal process:
\begin{equation}\label{eqdx}
\delta_sx(t)=\epsilon_s(t)e^{\omega_s(t)},
\end{equation}

where $\epsilon_s(t)$  and $\omega_s(t)$ are normal processes with
zero mean and standard deviations  equal to  $\sigma(s )$ and $\lambda(s)$ respectively.

Based on Eq.(\ref{eqdx}), we can write a probability density function for  $\delta_sx(t)$ as:

\begin{equation}\label{eqpdf}
P_{s}(\delta_s x)=\int _{0} ^{\infty}G_s(\ln \sigma(s)) \frac{1}
{\sigma(s)} F_s\left(\frac {\delta _{s}x} {\sigma(s)}\right) d(\ln
\sigma(s)),
\end{equation}
where
\begin{eqnarray}\label{eqpdf1}
G_s(\ln \sigma(s))&=&\frac {1} {\sqrt{2\pi} \lambda(s)}
\exp\left[-\frac
{(\ln\sigma(s)+\lambda^{2}(s))^2}{2\lambda^{2}(s)}\right],
\end{eqnarray}
and
\begin{eqnarray}\label{eqpdf2}
F_s\left(\frac {\delta _{s}x} {\sigma(s)}\right)&=&\frac {1}
{\sqrt{2\pi}} \exp\left[-\frac {\delta_{s}x^{2}}
{2\sigma^{2}(s)}\right].
\end{eqnarray}

since 
$\lambda^2(s)$ is the variance of the log-normal part of the process.
This parameter is the representative measure  of the non-Gaussianity of the
process; if  $\lambda^2(s)\rightarrow 0$, the PDF of $\delta_sx(t)$
converges to a Gaussian distribution. An estimation of $\lambda^2(s)$
versus a  scale $s$  is our way for presenting the effect  of large
fluctuations over different scales. Furthermore, for showing the  effects  of the rare events (in the PDF tails), high order moments of fluctuations of order
$q$, denoted by $m_{q}$,  can be calculated
\begin{equation}\label{eqm}
m_{q}(\delta_{s}x)=\langle
|(\delta_{s}x)|^{q}\rangle=\left[\int|\delta_{s}x|^{q}P_{s}
(\delta_sx)d(\delta_sx)\right]^{\frac{1}{q}}.
\end{equation}
A  large value of $m_{q}$ implies a  significant role of
rare events. If this   is  found to be independent of  $q$,  the process is called monofractal;  otherwise, it  is called a multifractal \cite{Shayeganfar}.

\subsection{Joint Multifractal Approach: the Bi-MRW method}

Muzy et al.  \cite{Muzy01,Muzy02} have proposed  the  bivariate Multifractal Random Walk (Bi-MRW) method for analyzing two non-Gaussian stochastic processes (${\bf
x}(t)=\{x_{1}(t),x_{2}(t) \}$)  simultaneously, when the increments of each time series  are supposed to be generated by the product of   normal and log-normal processes:
\begin{equation}
\delta_{s}\textbf{x}(t)=\left(\delta_{s}x_{1}(t),\delta_{s}x_{2}(t)\right)=\left(\epsilon_1^{(s)}(t)
e^{\omega_{1}^{(s)}(t)},\epsilon_{2}^{(s)}(t)e^{\omega_{2}^{(s)}(t)}\right),
\end{equation}
in which $(\epsilon_1^{(s)},\epsilon_2^{(s)})$ and
$(\omega_1^{(s)},\omega_2^{(s)})$ have joint normal PDF with zero
mean.

 Muzy {\it et al.},  generalizing  the MRW approach,  consider the cross-correlation of
stochastic variances of two processes  \cite{Muzy01}. Practically, 
$(\epsilon_1^{(s)},\epsilon_2^{(s)})$ have a covariance matrix  

\begin{equation}
\mathbf{\Sigma}_{(s)}\equiv\left(\begin{array}{cc}
\sigma_1^{2}(s)& \Sigma(s)\\
\Sigma_(s) & \sigma_2^{2}(s)\; \\
\end{array}\right)
\end{equation}
 where
$\Sigma(s)=\rho_{\epsilon}(s)\sigma_1(s)\sigma_2(s)$ \cite{Muzy01}.

The  covariance matrix of $(\omega_1^{(s)},\omega_2^{(s)})$ is
denoted by  $\mathbf{\Lambda}_{(s)}$ and  called a  multifractal matrix \cite{Muzy01}; it is   given by

\begin{equation}
\mathbf{\Lambda}_{(s)}\equiv\left(\begin{array}{cc}
\lambda_1^{2}(s) & \Lambda(s) \\
\Lambda(s) & \lambda_2^{2}(s)\; \\
\end{array}\right),
\end{equation}
where $\Lambda(s)=\rho_{\omega}(s)\lambda_1(s)\lambda_2(s)$ and
$\rho_{\omega}(s)$ is  the multifractal correlation coefficient. The PDFs
of $(\epsilon_1^{(s)}\;, \epsilon_2^{(s)})$, and $(\omega_1^{(s)}\;,
\omega_2^{(s)})$ have the following form:
\begin{eqnarray}\label{eq15}
F_s(\epsilon_1^{(s)},\epsilon_2^{(s)})&=&\frac{1}{2\pi\sqrt{{\rm
Det} (\mathbf{\Sigma}_{(s)})}}\
\exp\left({-\frac{\epsilon_{(s)}^{\rm
T}\cdot\mathbf{\Sigma}_{(s)}^{-1}\cdot\epsilon_{(s)}}{2}}
\right) \nonumber\\
\end{eqnarray}
\begin{eqnarray}\label{eq151}
G_s(\omega_1^{(s)},\omega_2^{(s)})&=&\frac{1}{2\pi\sqrt{{\rm
Det}(\mathbf{\Lambda}_{(s)})}}\
\exp\left({-\frac{\mathbf{\omega}^{\rm
T}_{(s)}\cdot\mathbf{\Lambda}^{-1}_{(s)}\cdot
\mathbf{\omega}_{(s)}}{2}}\right)\nonumber\;.\\
\end{eqnarray}
Therefore, the joint PDF of the fluctuations increment vector
$(\delta_{s}x_1,\delta_{s}x_2)$ is given by
\begin{widetext}
\begin{equation}\label{eq16}
P_s(\delta_{s}x_1,\delta_{s}x_2)=\int d(\ln\sigma_1(s))\int
d(\ln\sigma_2(s))G_s
\left(\ln\sigma_1(s),\ln\sigma_2(s)\right)\frac{1}{\sigma_1(s)}\frac{1}{\sigma_2(s)}
F\left(\frac{\delta_{s}x_1}{\sigma_1(s)},\frac{\delta_s
x_2}{\sigma_2(s)}\right)\;.
\end{equation}
\end{widetext}

It follows from the  above definitions  of  $G_s(\ln\sigma_1(s),\ln\sigma_2(s))$
and $F_s\left(\frac{\delta_s x_1}{\sigma_1(s)},\frac{\delta_s
x_2}{\sigma_2(s)}\right)$ 
that 

\begin{eqnarray}
&G_{s}\left(\ln \sigma_1(s),\ln\sigma_2(s)\right)=\frac{1}{2\pi
\sqrt{\lambda^2 _1(s)\lambda^2 _2(s)-\Lambda^2 (s)}}&\nonumber\\
&\exp\left[-\frac{\lambda_2^2(s)\left[\ln\sigma_1(s)+\lambda_1^2(s)\right]^2
+\lambda_1^2(s)\left[\ln\sigma_2(s)+\lambda_2^2(s)\right]^2-2\Lambda(s)
\left[\ln\sigma_1(s)+\lambda_1^2(s)\right]\left[\ln\sigma_2(s)+\lambda_2^2(s)\right]}{2\left(\lambda^2
_1(s)\lambda^2 _2(s)-\Lambda^2
(s)\right)}\right]&\nonumber\\
\end{eqnarray}
and

\begin{eqnarray}\label{fjoint}
&F_{s}\left(\frac {\delta_{s} x_{1}}
{\sigma_{1}(s)},\frac{\delta_{s} x_{2}}{\sigma_{2}(s)}\right)=\frac
{1} {2\pi\sqrt{\sigma^2 _1(s)\sigma^2 _2(s)-\Sigma^2
(s)}}\exp\left[-\frac{\left[\sigma_{2}(s)\delta_{s}
x_{1}\right]^2+\left[\sigma_{1}(s)\delta_{s} x_{2}\right]^2-
2\Sigma(s)\delta_sx_1\delta_sx_2}{2\left(\sigma^2 _1(s)\sigma^2
_2(s)-\Sigma^2
(s)\right)}\right].&\nonumber\\
\end{eqnarray}

It can be observed that $P_{s}(\delta_{s}x_1,\delta_{s}x_2)$  becomes equal to  $P_s(\delta_s x_1)
P_s(\delta_s x_2)$ when $\Lambda(s)$ and $\Sigma(s)$  tend to zero.

The $q$-th order moment of fluctuations increments for such two processes at scale $s$ can be written as
\begin{equation}\label{eqjm}
m^{joint}_{q}(\delta_{s}x_1,\delta_{s}x_2)=\langle
|(\delta_{s}x_1)|^{q}|(\delta_{s}x_2)|^{q}\rangle 
 =  \left[\int\int|\delta_{s}x_1|^{q}|\delta_{s}x_2|^{q}P_{s}
(\delta_sx_1,\delta_sx_2)d(\delta_sx_1)d(\delta_s
x_2)\right]^{\frac{1}{q}}.
\end{equation}


\section{Results}

\begin{figure}[ht]
\begin{center}
\includegraphics[width=0.6\linewidth]{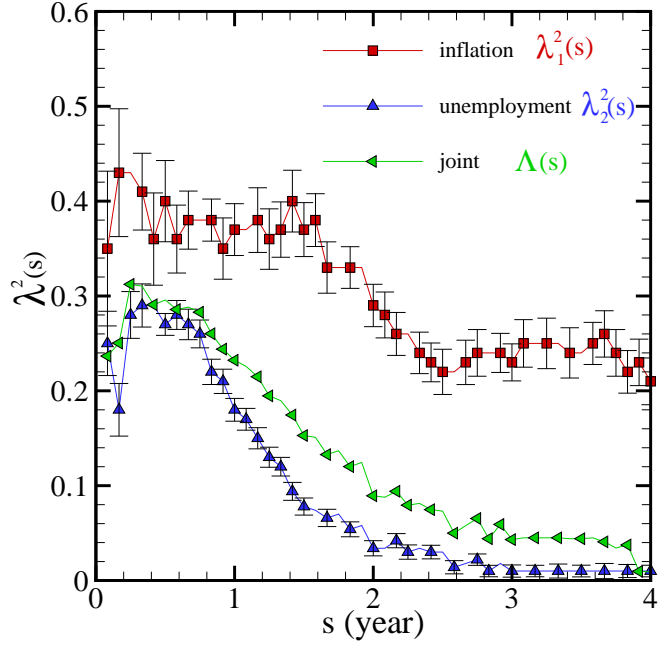}
\caption{
$\lambda^2(s)$ as a measure of non-Gaussianity for either unemployment "($\triangle$)"  or  inflation "$\Box$"  and the value of the multifractal coefficient $\Lambda(s)$ over different time scales "$s$". 
}\label{fig1}
\end{center}
\end{figure}

Thereafter, we can analyze both  inflation and unemployment rates, 
provided by the
U.S. Inflation Calculator \cite{inflation} and U.S. Bureau of Labor Statistics \cite{USBLBunemployment},  respectively. The data has been recorded monthly from January $1948$ until October $2018$. The non-Gaussian parameter
$\lambda^2(s)$ and the joint multifractal coefficient $\Lambda(s)$
at scale $s$ are obtained from the integral form of cascading rules
in Eqs. (\ref{eqpdf}) and (\ref{eq16}). The best values
for $\lambda^2(s)$ and $\Lambda(s)$ at scale $s$,  are for  the global minimum of the chi-square,
$\chi^{2}$ \cite{Agostini003,Sivia006}:
\begin{equation}
\chi^{2}(\Lambda(s);\Sigma(s))=\sum_{\delta_{s}{\bf
x}}\frac{\left[P_{\rm data}(\delta_{s}{\bf x})-P_{\rm
theory}(\delta_{s}{\bf
x};\Lambda(s),\Sigma(s))\right]^{2}}{\sigma^{2}_{\rm
data}(\delta_{s}{\bf x})+\sigma^{2}_{\rm theory}(\delta_{s}{\bf x};
\Lambda(s),\Sigma(s))},
\end{equation}
where $P_{\rm data}(\delta_{s}{\bf x})$ is the  joint PDF computed from
data, while  $P_{\rm theory}(\delta_{s}{\bf x};\Lambda(s),\Sigma(s))$ is the  theoretical 
joint PDF proposed in Eq.(\ref{eq16}).  By definition,  $\sigma^{2}_{\rm data}(\delta_{s}{\bf x})$ and $\sigma^{2}_{\rm
theory}(\delta_{s}{\bf x};\Lambda(s),\Sigma(s))$ are the mean standard deviation of $P_{\rm data}(\delta_{s}{\bf x})$ and $P_{\rm
theory}(\delta_{s}{\bf x};\Lambda(s),\Sigma(s))$, respectively.

The best value of  $\Lambda(s)$ for  the theoretical joint PDF is obtained from the fit of the joint PDF to the data:

\begin{eqnarray}\label{eq7}
\chi^{2}(\Lambda(s))&=&\sum_{\delta_{s}{\bf x}}\int d\Sigma(s)
\left(\frac{\left[P_{\rm data}(\delta_{s}{\bf x},s)-P_{\rm
theory}(\delta_{s}{\bf
x};\Lambda(s),\Sigma(s))\right]^{2}}{\sigma^{2}_{\rm
data}(\delta_{s}{\bf x},s)+\sigma^{2}_{\rm theory}(\delta_{s}{\bf
x};
\Lambda(s),\Sigma(s))}\right),\nonumber\\
\end{eqnarray}

The parameter $\lambda^2(s)$ is depicted for
inflation, $\lambda_1 ^2(s)$, and unemployment, $\lambda_2 ^2(s)$,  in Fig. \ref{fig1}.
It is seen  that $\lambda_1 ^2(s)$ is large at all scales, whereas
$\lambda_2 ^2(s)$ is large at scales smaller than one year. Large
values of $\lambda_1 ^2(s)$ imply that rare events occurring in the
inflation  rate make its PDF  non-Gaussian.  For unemployment,  $\lambda_2
^2(s)$ tends to zero at scales larger than one  year. This scaling
dependency of $\lambda_2 ^2(s)$ implies that the occurrence of rare
events in unemployment provides  a non-Gaussian behavior at short
time scales,  but after one year it tends to  a Gaussian state.

Concerning the joint multifractal coefficient $\Lambda(s)$,  see  Fig. \ref{fig1}, we observe  that it has its highest values for  scales lower than one year.  Thereafter, $\Lambda(s)$ decreases relatively fast over the scales below two years.
This  is compatible with some beliefs about the effect of inflation on the joint relation in the short run and its ineffectiveness in the long-run \cite{ball}.

\begin{figure}
    \begin{center}
        \includegraphics[width=5.3cm,height=5cm,angle=0]{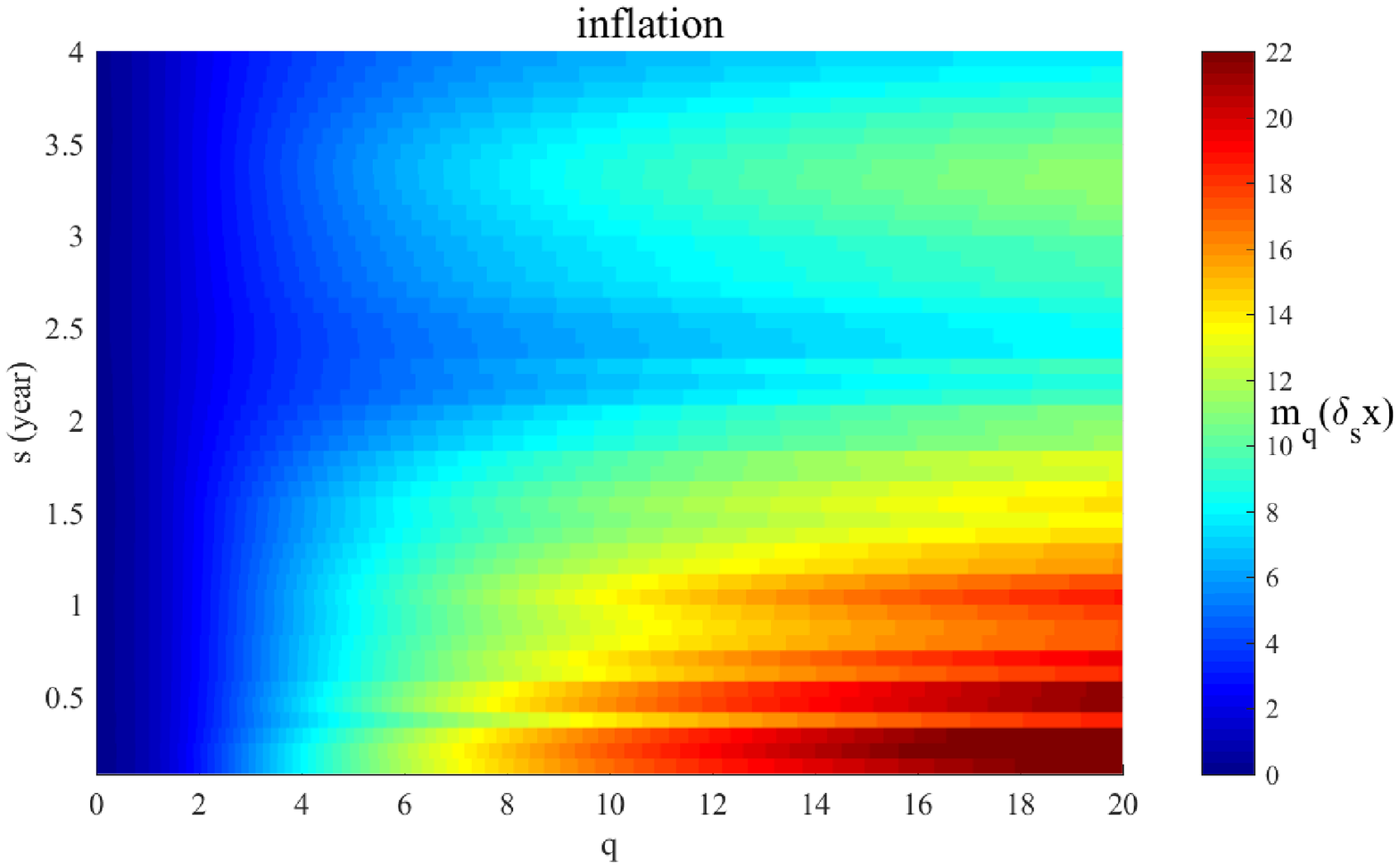}
        \includegraphics[width=5.3cm,height=5cm,angle=0]{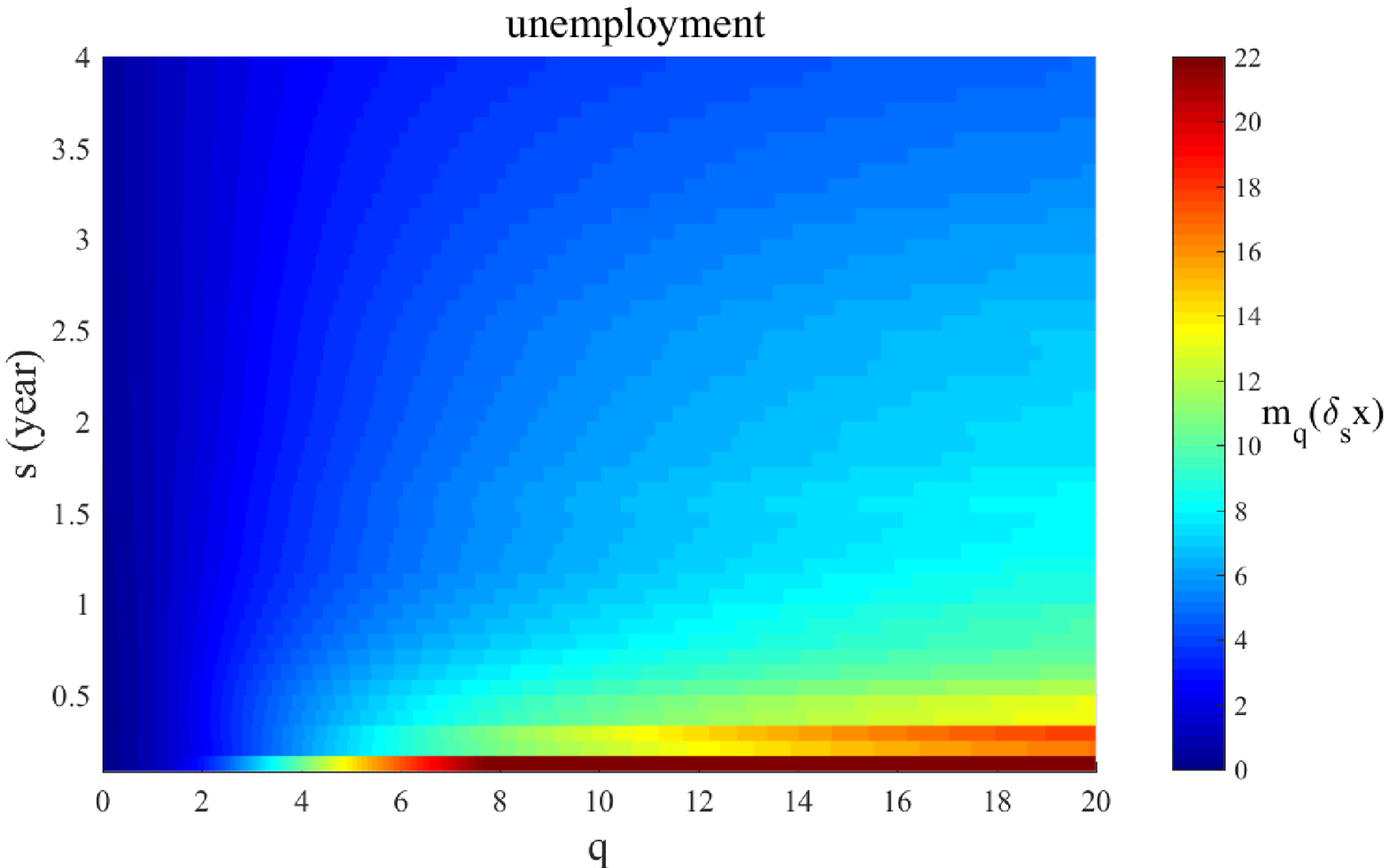}
        \includegraphics[width=5.3cm,height=5cm,angle=0]{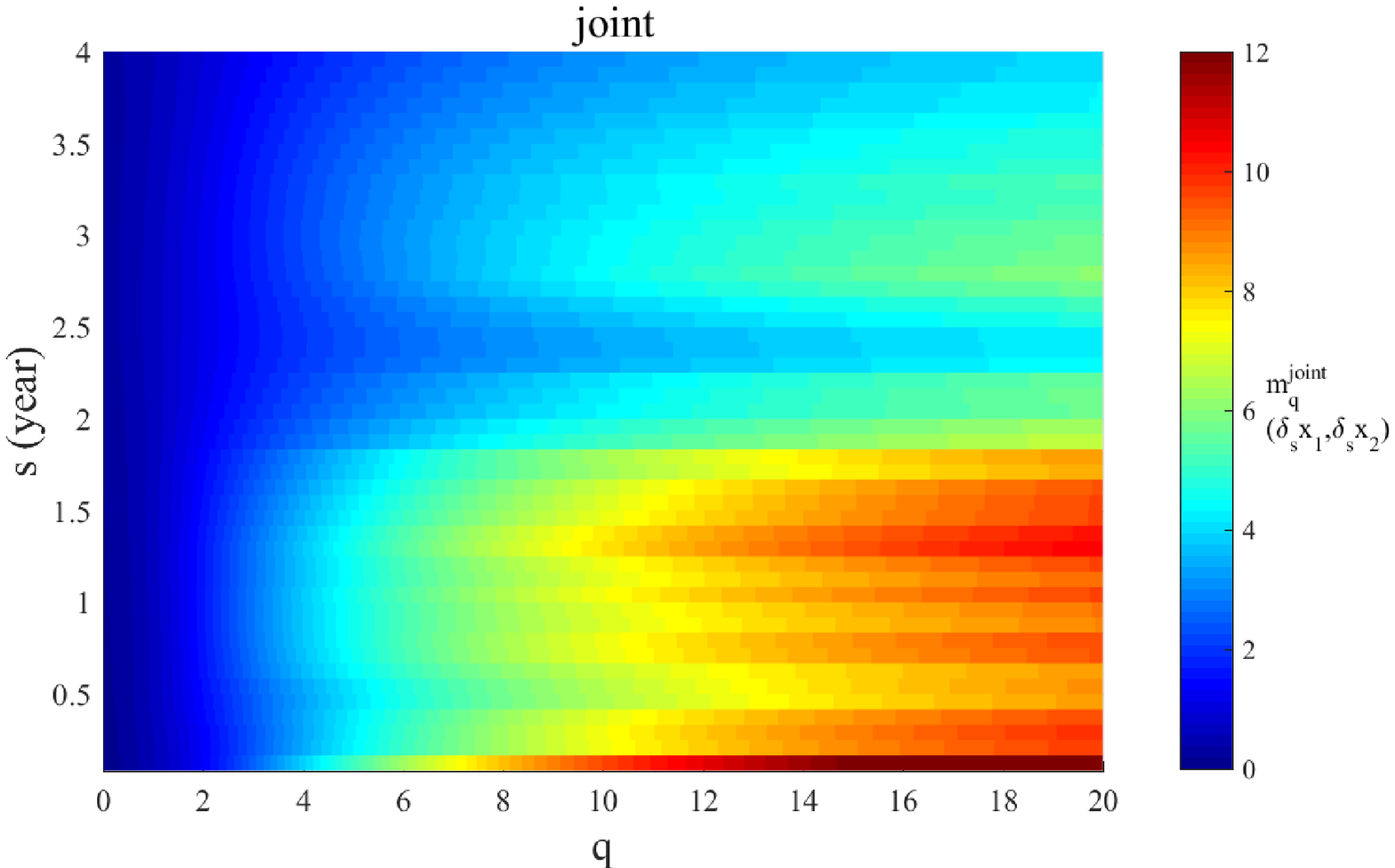}
    \caption{The colour map of the moment $m_q(\delta_s x)$ (from Eq. (\ref{eqm})) and joint moment $m^{joint}_{q}(\delta_{s}x_1,\delta_{s}x_2)$ (from Eq. (\ref{eqjm})) for different values of $q$ and different time scale $s$. As it can be seen, for unemployment  (middle figure) the value of high  moments drop rapidly for scales above six months. }
    \label{fig2}
    \end{center}
\end{figure}

To improve our understanding about the behavior of the rare events,  higher moments of the variables'  increments have been measured  for various orders and various  time scales. Recall that high  moments are more influenced by the rare events in the tails of the PDF.

 In Fig. \ref{fig2},
color intensity plot of such high moments  are depicted  for different time scales.
As it can be seen, the value of the high  moments of the unemployment  rates have their largest values for  scales below six months. Above six months the moments drop rapidly. In contrast to the unemployment case, the value of high moments for the inflation rates is relatively noticeable for all scales but is higher in at scales below $2$ years.

The right panel in Fig.\ref{fig2} illustrates that the behavior of the joint moment is more similar to the inflation case where a  noticeable reduction can be observed for scales above $2$ years. This means that large fluctuations in inflation and unemployment are  more strongly  coupled  above this time interval.

\begin{figure}[h]
    \begin{center}
        \includegraphics[width=5.3cm,height=5cm,angle=0]{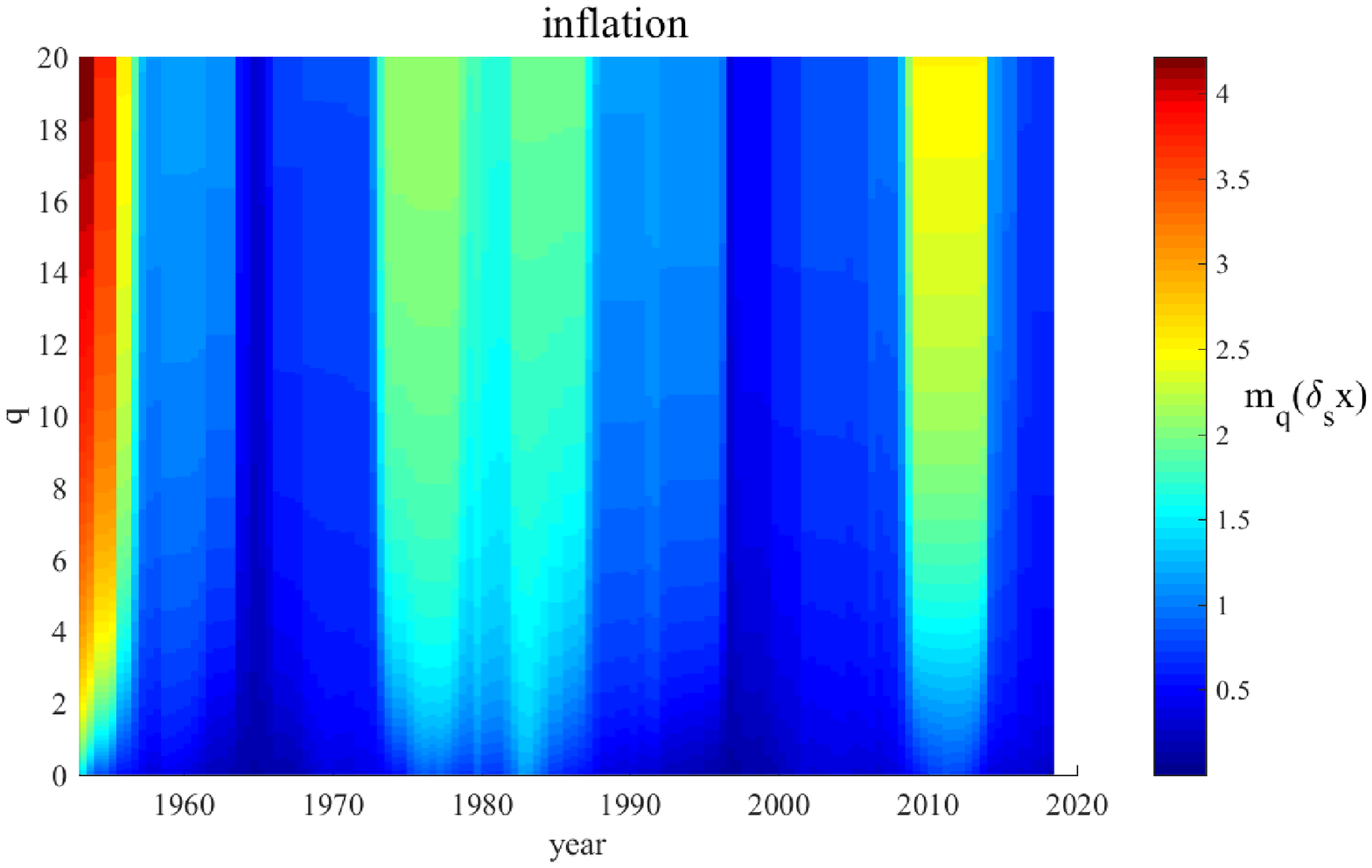}
        \includegraphics[width=5.3cm,height=5cm,angle=0]{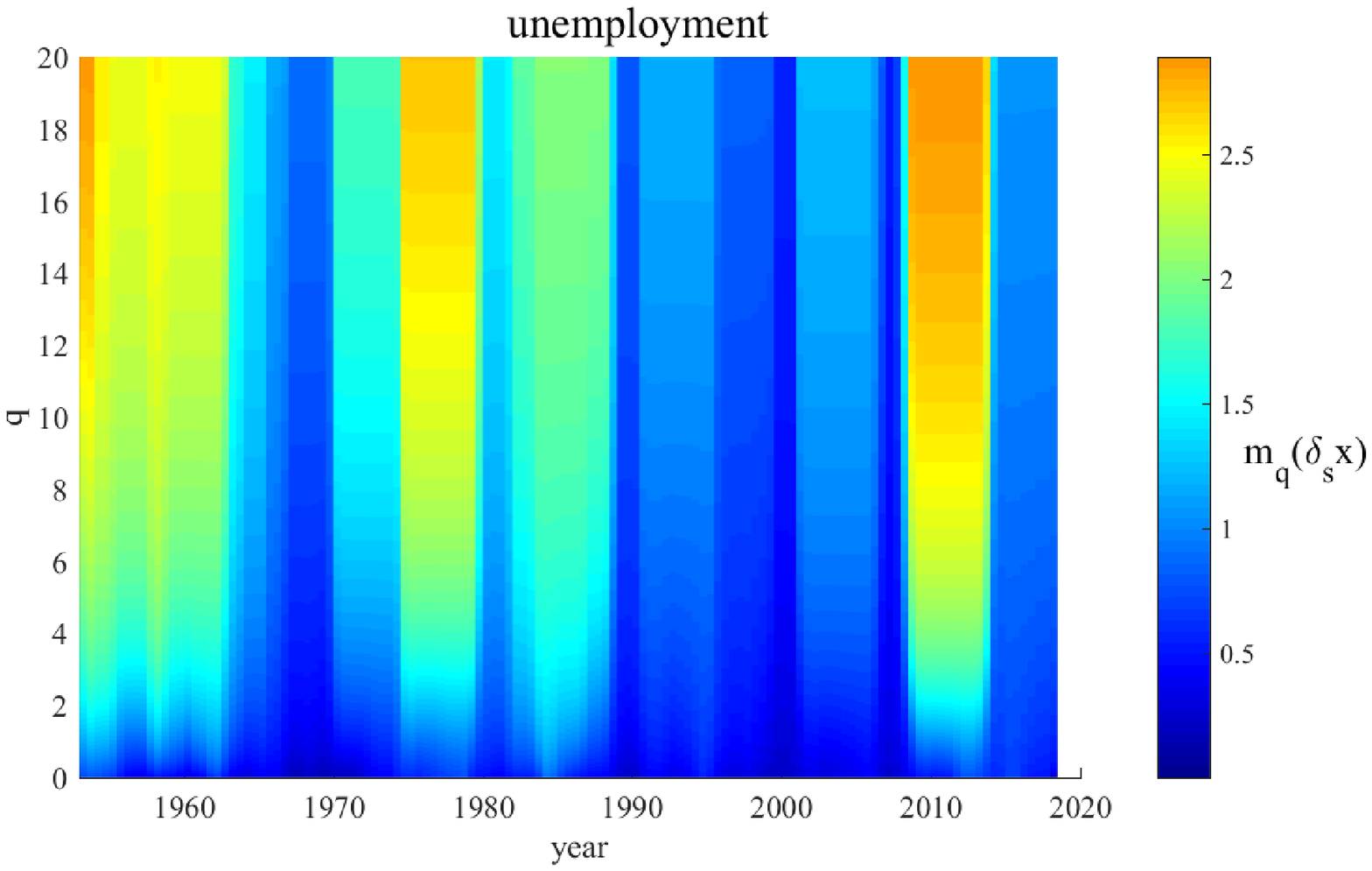}
        \includegraphics[width=5.3cm,height=5cm,angle=0]{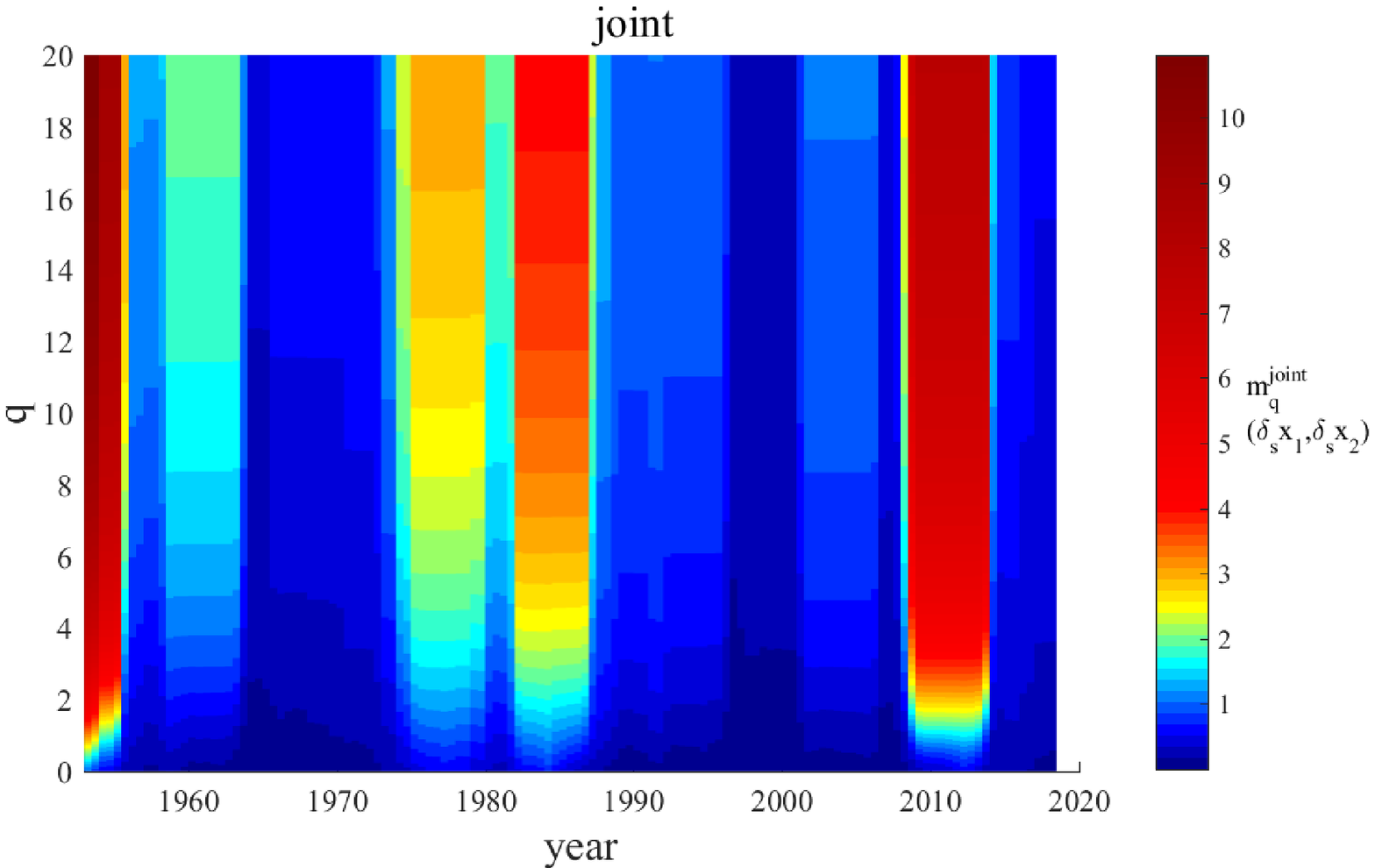}
    \caption{The moments $m_q(\delta_s x)$ and $m^{joint}_{q}(\delta_{s}x_1,\delta_{s}x_2)$ of order $q$ are estimated at scale 1 year from $1948$ to $2018$ for inflation (left), unemployment (middle) and the pair  of variables (right).}
    \label{fig3}
    \end{center}
\end{figure}

At the next step, the behavior of high moments of inflation and unemployment   are investigated as time evolved between   January $1948$ and October $2018$. A window of five years is chosen and the moments $m_q(\delta_s x)$  calculated  for the  one year  scale ($s=1$). The  window is slowly moved along the time axis. Results have been depicted in Fig. \ref{fig3}.

As it can be seen, from the red bands, the joint relation has sharply grown  at    critical periods in the modern  history of the US economy, i.e. the volatile postwar period, the stagflation of $70_{th}$, the period of Volcker deflationary program, in the 80's, and the great  recent recession of 2008-2009.

If  the inflation rate likely dominates the  coupled  relation at the post WWII time, in contrast, the unemployment rate seems to dominate the joint relation in both other cases.


\section{Conclusion}

 Inflation and unemployment are dependent variables with non-Gaussian PDFs; however, the scale and intensity of this dependency and their coupling effects  have been, and are still, much debated. The controversy finds its importance when the government aims to impose an expansionary monetary policy over the economic crises.
 Many researchers have discussed the relation between inflation and unemployment; recall one  Friedman \cite{friedman1977nobel}, among others,
 \cite{rochon2018relationship}.
 In this work, we were interested in large and rare fluctuations in the measures of these economic variables, in order to  observe possible enhancements in themselves and in  their coupling. We have focused on the non-Gaussianity of the signals' PDFs and their coupling.
The Multifractal Random Walk (MRW) is known as a good approach that detects the non-Gaussianity of PDFs  through the parameter $\lambda^2(s)$, the variance of the log-normal process. Moreover,
the  bivariate Multifractal Random Walk (Bi-MRW) method is useful for analyzing two non-Gaussian stochastic processes through the corresponding variance   $\Lambda^2(s)$.

By analysing  70 years of US data via these techniques, the non-Gaussianity of the PDFs of unemployment and inflation and also their joint relation have been detected. The non-Gaussianity parameter  $\lambda^2(s)$ of the unemployment  rate data is smaller than that  for the inflation and  that for the  coupled  relation  $\Lambda^2(s)$. It is observed that after one year, the behavior of this parameter for the unemployment case tends toward a normal condition,  but  on the contrary, for the inflation data, the non-Gaussianity parameter persists for  all studied scales.   The non-Gaussianity of the joint relation  decreases to small values for  scales above $2$ years.

Also, this behavior is observable in the high-order moments of these (three) variables.
Moving  the observation window over time, it is discovered  that the non-Gaussianities of the parameters and of  their coupling substantially grow over the critical periods of the economy.

\end{document}